\documentclass[12pt]{article}

\usepackage{natbib}

\newcommand{\disp}{\displaystyle}
\newcommand{\ti}{\tiny}

\newcommand{\xsat}{$X_{i,{\rm sat}}$}
\newcommand{\xliq}{$X_{\rm liq}^{\rm ini}$}
\newcommand{\xsol}{$X_{\rm sol}^{\rm ini}$}
\newcommand{\xevap}{$X_{\rm evap}^{\rm fin}$}

\newcommand{\REVfirst}[1]{{#1}}

\newcommand{\REVsec}[1]{{#1}}
\newcommand{\REVsecFINAL}[1]{{#1}}

\newcommand{\variouscom}[1]{{#1}}
\begin{document}

\title{On the chemical composition of Titan's dry lakebed evaporites}
\begin{center}
{\bf\Large On the chemical composition of Titan's dry lakebed evaporites}
\end{center}

\vspace{0.5cm}
\begin{itemize}
  \item[] {\small \textbf{Cordier}, D., Universit{\'e} de Franche-Comt{\'e}, Institut UTINAM, \\CNRS/INSU, UMR 6213, 25030 Besan\c{c}on Cedex, France.}
  \item[] {\small \textbf{Barnes}, J. W., Department of Physics, \\University of Idaho, Engineering-Physics Building,\\ Moscow, ID 83844, USA.}
  \item[] {\small \textbf{Ferreira}, A., Departamento de Engenharia Quimica,\\ Universidade de Coimbra, Coimbra 3030-290, Portugal.}
\end{itemize}
\vspace{0.5cm}

\begin{minipage}{12cm}%
                 \textbf{Abstract} {\small   
  {Titan, the main satellite of Saturn, has an active cycle of methane in its troposphere.
   Among other evidence for a mechanism of evaporation at work on the ground, dry lakebeds have been
   discovered. Recent \emph{Cassini} infrared observations of these empty lakes have revealed a surface composition
   poor in water ice compared to that of the surrounding terrains --- suggesting the existence of organic evaporites deposits.
   The chemical composition of these possible evaporites is unknown.
   }
   {In this paper, we study evaporite composition}
   {using a model that treats both organic solids dissolution and
   solvent evaporation.}
   {Our results suggest the possibility of large abundances of butane and acetylene in the lake evaporites. 
	However, due to
   uncertainties of the employed theory, these determinations have to be confirmed by laboratory experiments.}}
\end{minipage}
                 
\section{Introduction}
\label{sec:intro}

 For a long time the existence of liquid hydrocarbons at the surface of Titan has been suspected 
\citep[][]{sagan_dermott_1982,lunine_etal_1983,lunine_1993a,lunine_1993b}. The dark features observed by
\cite{stofan_etal_2007} in the north polar region were the 
\REVsec{first confirmed lakes or seas of hydrocarbons}.
 Subsequently, other evidence for the RADAR-dark areas' lacustrine nature was 
found in the RADAR and IR ranges, to the extent that the existence of lakes/seas is now rather well established.
In fact, the number of detected manifestations \citep[e.g.][]{turtle_eal_2011a,turtle_eal_2011b} of an active tropospheric 
methane hydrologic cycle is increasing. The lakes are expected to take part 
\REVsec{in this cycle},
providing methane and/or ethane to the
atmosphere through evaporation processes.\\
   In past years, the signature of lake evaporation has been actively researched. Already, \cite{stofan_etal_2007} noticed 
features showing margins similar to those of established lakes but having a RADAR surface backscatter similar to the surrounding 
terrain, suggesting the occurrence of an evaporation process in the recent past. \cite{barnes_etal_2009a} performed a detailed study 
of shoreline features of Ontario Lacus, the largest southern latitude 
\REVsec{lake}.
These authors interpreted the 5-$\mu$m bright annulus 
around Ontario Lacus as a dry, low-water ice content zone, possibly corresponding to a deposit of fine-grained organic condensates. 
These patterns, created by the shoreline recession, could have been caused by an evaporation episode. In their study of the same system, 
\cite{wall_etal_2010} reported evidences for active shoreline processes.  Although evidence for short-term changes in the extent of
Ontario Lacus has been put forward \citep{turtle_eal_2011b}, a subsequent reanalysis came to the conclusion that there is no indication 
of lake extent changes in the \emph{Cassini} dataset \citep{cornet_etal_2012}.
\cite{hayes_etal_2011} noticed that some observed dry lakebeds in Titan's arctic appear to be brighter than their exteriors in both
nadir and off-nadir observations, which suggests compositional  differences. However \cite{hayes_etal_2011} were not able to exclude the
possibility of an infiltration of liquids into a subsurface hydrologic system. \cite{barnes_etal_2011} used a sample of several lakes and
lakebeds located in a region south of the Ligeia Mare. They obtained a strong correlation between RADAR-empty lakes and 5-$\mu$m-bright unit
interpreted as low-water ice content areas.\\
  As mentioned by \cite{barnes_etal_2011} these observed dry lake floors cannot be made only of sediments, indeed a pure sedimentary 
origin of these deposits would produce lakebed showing a 5-$\mu$m-brightness similar to that of their surrounding zones. One 
possible explanation proposed by \cite{barnes_etal_2011} consists of evaporation of the solvent (here a mixture of methane
and ethane) 
\REVsec{yielding}
to the saturation of the dissolved solutes. The top layer of the resulting evaporites is being 
observed now in dry lakebeds if this idea is correct.
This paper is devoted to an exploration of the evaporite scenario on the theoretical side. We have developed a model allowing for the 
computation of the chemical composition of such evaporites.\\
  This paper is organized as follows. In Sect.~\ref{model}, we outline our model for calculating chemical composition of
putative evaporite deposits in dry lakebeds. Sect.~\ref{compu} is devoted to evaporite composition computations, and we discuss 
our results and conclude in Sect.~\ref{discuss}.

\section{\label{model}Model Description}

   We consider a portion of a 
\REVsec{Titan lake}
of uniform depth $h$ that has a free surface of area $S$ in contact with 
the atmosphere. 
For the sake of simplicity, methane, ethane and nitrogen are considered to be the only volatile compounds; they form a ternary
mixture which will be our solvent. The presence of H$_{2}$, Ar and CO is neglected as they have low abundances 
\REVsec{in the atmosphere and as a consequence in the solution}; 
C$_{3}$H$_{8}$ and 
C$_{4}$H$_{8}$ are also not taken into account because C$_{2}$H$_{6}$ seems to be much more abundant and their behaviors
should not be very different than that of ethane.
  In addition to the solvent chemistry itself, \REVfirst{we considered species in the solid state under Titan's surface
thermodynamic conditions that may dissolve in the solvent. In the following, for short, we will simply call these compounds
``dissolved solids" or ``solutes": they include all the species, except those belonging to the solvent (\emph{i.e.}, methane, ethane and nitrogen). These} 
supposed dissolved \REVfirst{species} are ultimately the products of the complex photochemistry taking place in
the upper Titan atmosphere. In this work, we used the same list of solid compounds as in previous papers
\citep[][hereafter respectively C09, C10 and C12.  Note that \cite{cordier_etal_2013a} is an erratum of 
\cite{cordier_etal_2009}.]{cordier_etal_2009,cordier_etal_2010,cordier_etal_2012} 
%
These species appear to be among the
main products found by photochemical 1D-models of \cite{lavvas_etal_2008a,lavvas_etal_2008b} and \cite{vuitton_etal_2008}; the list is
displayed in Table~\ref{solids}. 
\REVsecFINAL{This list differs from the list of species detected by CIRS\footnote{Composite Infrared Spectrometer, an instrument onboard the
Cassini spacecraft.} 
\citep[\emph{i.e.}, CH$_4$, C$_2$H$_2$, C$_2$H$_4$, C$_2$H$_6$, C$_3$H$_8$, CH$_3$C$_2$H, 
C$_4$H$_2$, C$_6$H$_6$, HCN, HC$_3$N, C$_2$N$_2$, CO, CO$_2$ and H$_2$O, see][]{vinatier_etal_2010a}
as CIRS observations relate to Titan's stratosphere and do not
imply that these species reach the thermodynamic conditions of their precipitation to the ground. In addition, some species 
\citep[for instance C$_4$H$_3$, C$_4$H$_4$, C$_4$H$_5$; see Tab.~5 of ][]{lavvas_etal_2008a}
are included in models while they are not yet observed by Cassini instruments.}
Thanks to their melting temperatures ranging between 136.0 K (C$_{4}$H$_{10}$) and 279.1 K (C$_{6}$H$_{6}$),
all these molecules are in solid state under the Titan's surface conditions  
\citep[$T \sim 90$ K in the region of lakes;][]{jennings_etal_2009}. Although it has still to be confirmed by laboratory experiments, 
the materials listed in Table~\ref{solids} are theoretically predicted to be soluble in a mixture of methane and ethane. 
Indeed, the Hildebrand's  solubility parameters $\delta$'s for these solids
\citep[see][]{poling_2007,ahuja_2009} are close to the ethane value (see Table~\ref{solids}).
   We implement a numerical calculation for the dynamic composition evolution of liquid mixtures using discrete timesteps.
 At each time $t$, the saturation mole fraction $X_{i,{\rm sat}}$ of each dissolved solid species $i$ is computed via 

\begin{equation}\label{Xs}
   \mathrm{ln} \, \Gamma_{i} X_{i,{\rm sat}} = - \frac{\disp\Delta H_{i,{\rm m}}}{\disp R T_{i,{\rm m}}} \,
       \left(\frac{T_{i,{\rm m}}}{T} - 1\right)
\end{equation}

  This relation can be found, for instance, in the Section~8-16 of the textbook by \cite{poling_2007} (hereafter POL07). 
The physical significance of Equation (\ref{Xs}) is the existence of a thermodynamic equilibrium between the considered precipitated 
solid $i$ and the liquid solution. The enthalpy of melting is denoted $\Delta H_{i, {\rm m}}$, whereas $T$ and $T_{i, {\rm m}}$ are 
respectively the current temperature of the lake and the melting temperature of molecule $i$; $R$ is the constant of ideal gases and $\Gamma_{i}$ is the
activity coefficient. 
Although Equation (\ref{Xs}) has previously been used in several published works \citep[][hereafter D89, and also C09, C10 
and C12]{dubouloz_etal_1989}, we recall that it is approximate and its validity will be discussed in Section~\ref{discuss}.

As the thermodynamic computations in the frame of the regular solution theory are uncertain (see C12) due to the lack of knowledge of
thermodynamic data, we have distinguished two cases: the approximation of the ideal solution for which all
the $\Gamma_{i}$'s are equal 
\REVsec{to unity}, 
and the non-ideal regular solution. In the case of an ideal solution, 
the molecules of the same species and those of different species interact with same intensity. 
For a non-ideal solution model, the $\Gamma_{i}$'s are computed in the frame 
\REVsec{of regular solution theory}
(see D89, C09, POL07) 
\REVfirst{in which the intermolecular interactions of involved species are such that the resulting entropy of mixing is equal 
to that of an ideal solution with the same composition: zero excess  entropy, with the volume of mixing at zero. However in 
contrast with ideal solutions, regular solutions have mixing enthalpy with  nonzero values. The regular-solution theory provides 
a good and useful semiquantitative representation of real behavior for solution containing nonpolar components as is the case of 
the mixtures under study in this work and the results based on this theory are in general considerably improved over those 
calculated by Raoult's law. In the context of the mixtures studied here it is expected that even though results may not possess 
extreme accuracy they are also hardly ever very bad, providing a valuable guide for future work.}
We emphasize that the
$\Gamma_{i}$'s are 
\REVsec{functions} 
of $X_{i}$'s, a fact which leads to numerical complications.  The temperature $T$ of
the liquid remains unchanged during the whole evaporation process. Note, by the way, that for the temperatures relevant 
(see Table~\ref{solids} for melting temperatures values) in our context, the right-hand side of Equation (\ref{Xs}) is negative, 
leading to mole fractions lower than the unity, at least in the case of an  ideal solution.
\begin{table}[ht]
\caption[]{Solids assumed to be dissolved in the lake and some of their properties. The Hildebrand's solubity parameter $\delta$
has to be compared to the value for methane and ethane at the same temperature (i.e. 90 K), which are respectively 
$1.52 \times 10^{4}$ (J\,m$^{-3}$)$^{1/2}$ and $2.19 \times 10^{4}$ (J\,m$^{-3}$)$^{1/2}$. For comparison purposes, for H$_{2}$O
 $\delta \sim 5 \times 10^{4}$ (J\,m$^{-3}$)$^{1/2}$.}
\begin{center}
{\small
\begin{tabular}{lcccc}
\hline
\hline 
Species         & Precipitation                     & $\delta$                            & Melting     & Enthalpy             \\
                & rate                              &                                     & temperature & of melting           \\
                & {\ti molecules\,m$^{-2}$\,s$^{-1}$} & {\ti $10^{4}$ (J\,m$^{-3}$)$^{1/2}$} & {\ti(K)}    & {\ti(kJ\,mol$^{-1}$)} \\
HCN             & $1.3 \times 10^{8}$$^{(a)}$       & $2.99$                              & 260.0       &  $8.406$             \\
C$_{4}$H$_{10}$ & $5.4 \times 10^{7}$$^{(a)}$       & $1.91$                              & 136.0       &  $4.661$             \\
C$_{2}$H$_{2}$  & $5.1 \times 10^{7}$$^{(a)}$       & $2.28$                              & 192.4       &  $4.105$             \\
CH$_{3}$CN      & $4.4 \times 10^{6}$$^{(a)}$       & $2.92$                              & 229.3       &  $6.887$             \\
CO$_{2}$        & $1.3 \times 10^{6}$$^{(a)}$       & $1.98$                              & 216.6       &  $9.020$             \\
C$_{6}$H$_{6}$  & $1.0 \times 10^{6}$$^{(b)}$       & $2.48$                              & 279.1       &  $9.300$             \\
\hline
\end{tabular}
}
\\{\small$^{(a)}$\cite{lavvas_etal_2008a,lavvas_etal_2008b}; $^{(b)}$\cite{vuitton_etal_2008}.}
\end{center}
\label{solids}
\end{table}

  The equilibrium solid-solution written as Equation (\ref{Xs}) must be complemented by the Principle of Matter Conservation. If we 
denote the total 
\REVsec{number of moles of all species}
at time $t$ in the liquid by $N(t)$; at time $t+\mathrm{d}t$, one can write for a 
lake with surface area $S$
\begin{equation}\label{conserv}
\begin{array}{lcl}
   N(t+\mathrm{d}t) = N(t)  & - & F_{\rm CH_{4}} \times S \times \mathrm{d}t -  F_{\rm C_{2}H_{6}} \times S \times \mathrm{d}t
                                  - F_{\rm N_{2}} \times S \times \mathrm{d}t  \\
                            & - & \sum_{i,{\rm sat}, t+\mathrm{d}t} (X_{i}(t) \, N(t) - X_{i,{\rm sat}}^{\rm ideal}N(t+\mathrm{d}t))
\end{array}
\end{equation}

where $F_{\rm CH_{4}}$, $F_{\rm C_{2}H_{6}}$ and $F_{\rm N_{2}}$ are the assumed respective 
evaporation rates (in
mole\,m$^{-2}$\,s$^{-1}$) of CH$_{4}$, C$_{2}$H$_{6}$ and N$_{2}$. The terms containing the $F_{i}$'s represent the evaporation 
while the sum, which refers to species reaching the saturation at $t+\mathrm{d}t$, corresponds to matter that precipitates and deposits
on the lake floor.

The number of mole of species $i$ available in the volume $S \times H$ of lake, at time $t$, is denoted by $n_{i}(t)$. Thus
we arrive at the
simple relation $n_{i}(t)= X_{i} N(t)$. Our algorithm consists of several steps.  For the first one, we compute 
$N^{(0)}(t+\mathrm{d}t)$. This is an estimation of the total number of moles (in volume $S \times H$ of lake) remaining after the time step
$\mathrm{d}t$ during which only evaporation of methane, ethane and nitrogen is taken into account: 
\begin{equation}\label{N0}
N^{(0)}(t+\mathrm{d}t) = N(t) - (F_{\rm CH_{4}} + F_{\rm C_{2}H_{6}} + F_{\rm N_{2}}) \times S \times \mathrm{d}t\mathrm{~~~.}
\end{equation}

From this, we can infer the corresponding mole fraction 
\begin{equation}\label{X0}
X^{(0)}_{i} = \frac{\disp n_{i}(t+\mathrm{d}t)}{\disp N^{(0)}(t+\mathrm{d}t)}
\end{equation}

with $n_{i}(t+\mathrm{d}t)= n_{i}(t)$ for all species except those belonging to the solvent. If, as a first attempt, we work in the
frame of the ideal solution theory, then we have to compare the $X^{(0)}_{i}$'s to the mole fractions at saturation (for an ideal solution) 
$X^{\rm ideal}_{i,{\rm sat}}$. The molecules for which the criterion
is fulfilled are presumed to precipitate. Their abundances are fixed to the saturation value 

\begin{equation}
  X_{i}(t+\mathrm{d}t) = X_{i,{\rm sat}}^{\rm ideal} \;\; {\rm(only\; species\; that\; saturate)}~~~,
\end{equation} 

 and the total number of moles at $t +\mathrm{d}t$ in the volume $S \times H$ is recomputed to be
\begin{equation}\label{Ntpdtideal}
  N(t+\mathrm{d}t) = \frac{\disp \left[1-\sum_{\rm sat} X_{i,{\rm sat}}\right] N(t) - (F_{\rm CH_{4}} 
                     + F_{\rm C_{2}H_{6}} + F_{\rm N_{2}}  ) \times S 
                     \times \mathrm{d}t }{\disp 1 - \sum_{\rm sat} X_{i,{\rm sat}}^{\rm ideal}} ~~~~.
\end{equation}

  Finally, the abundances  at $t +\mathrm{d}t$ of species that do not saturate can be easily derived
\begin{equation}
  X_{i}(t+\mathrm{d}t) = \frac{\disp n_{i}(t)}{\disp N(t+\mathrm{d}t)}
\end{equation}

  For a given species $i$ that saturates, the rate of evaporite formation is then given by

\begin{equation}
  F_{i}^{\rm evap}(t+\mathrm{d}t) = \frac{\disp X_{i}(t) \, N(t) 
                       - X_{i,{\rm sat}}^{\rm ideal} \, N(t+\mathrm{d}t)}{\disp S \times \mathrm{d}t} ~~~~.
\end{equation}

 The chemical composition of the formed evaporite at time $t+\mathrm{d}t$ is given by 
\begin{equation}
      X_{i}^{\rm evap}(t+\mathrm{d}t) = 
         \frac{\disp F_{i}^{\rm evap}(t+\mathrm{d}t)}{\disp\sum_{j,{\rm sat}} F_{j}^{\rm evap}(t+\mathrm{d}t)} ~~~~.
\end{equation}

      When the solution is considered as a ``real'' solution, i.e. the activity coefficients are not taken to be equal to
the unity and depend on the mole fractions, then

\begin{equation}
    \Gamma_{i}(t+\mathrm{d}t) = \Gamma_{i}(X_{1}(t+\mathrm{d}t), X_{2}(t+\mathrm{d}t), ...) ~~~~.
\end{equation}

    The scheme allowing the computation of evaporite composition when the solution is non-ideal is rather
similar to that of the case of an ideal solution. However, there are some differences: the total number of moles 
$N(t+\mathrm{d}t)$ can no longer be computed
with Eq.~\ref{Ntpdtideal}. Instead, if $N_{\rm sat}$ is the number of species saturating at $t+\mathrm{d}t$, then we solve the
system of $(N_{\rm sat}+1)$ equations composed of Equation (\ref{conserv}) and the $N_{\rm sat}$ 
equations similar to Equation (\ref{Xs}). The unknowns are
$N(t+\mathrm{d}t)$ and the $N_{\rm sat}$ mole fractions at saturation $X^{\rm non-ideal}_{k,{\rm sat}}$. The resolution of that
non-linear system is performed using a multi-dimensional Newton-Raphson's method \citep[see][]{Num_Recipes}.

  Several authors \citep{mitri_etal_2007,tokano_2005,tokano_2009} have studied the dynamics of Titan's lakes. They have used the 
bulk aerodynamical model, introduced for Earth's climate model by \cite{fairall_elat_1996}. In this model the evaporation rate 
is given by

\begin{equation}
  E = \rho_{\rm air} \, K \, (q^{*} - q) \, u_{r}
\end{equation}

where $E$ is expressed in kg\,m$^{-2}$\,s$^{-1}$, $\rho_{\rm air}$ as the density of the air, $K$ is a transport coefficient (a purely
aerodynamic quantity), $q^{*}$ and $q$ are saturation specific humidity and specific humidity respectively, and $u_{r}$ is
the horizontal component of the averaged wind speed relative to the surface at a given height $z_{r}$. As we can see, the rate $E$
depends on $q$ and $u_{r}$, quantities that are supposed to undergo 
\REVsec{substantial}

variations during a Titan's year. Thus the rate $E$
is expected to experience significant variations during Titan's seasonal cycle. To obviate this problem we have 
chosen to use two 
definitions of time and two time-scales, respectively corresponding to the evaporation of CH$_{4}$ and the evaporation of 
C$_{2}$H$_{6}$. Methane is known to be much more volatile than ethane. An order of magnitude of the ratio
$\alpha_{\rm evap}= F_{\rm C_{2}H_{6}}/F_{\rm CH_{4}}$ can be evaluated thanks to the Hertz-Knudsen 
\cite[see][and references therein]{ward_fang_1999}: one finds that $\alpha_{\rm evap} \sim 10^{-4}$ for a 
temperature $T \simeq 90$~K.
In our model we fixed $\alpha_{\rm evap}$ to this value. The evaporation rate of nitrogen is 
scaled to the C$_{2}$H$_{6}$ one, in order to insure a N$_{2}$ content in the liquid 
\REVsec{preventing it from freezing} 
\citep[see][]{mitri_etal_2007}. Therefore, all the CH$_{4}$ content evaporates on a short time-scale $\tau_{\rm CH_{4}}$ during 
which the time $t_{\rm CH_{4}}$ is defined so that the evaporation rate $F_{\rm CH_{4}}$ is constant. We emphasize that this 
operation is equivalent to an implicit non-linear re-scaling of time. A similar definition
is adopted for $t_{\rm C_{2}H_{6}}$, the ``variable time'' valid during ethane evaporation, 
as $F_{\rm CH_{4}} \gg F_{\rm C_{2}H_{6}}$,
the quantity of ethane escaping the lake during methane evaporation, is negligible. This way, all the details of the possible
events (\emph{e.g.}, low or high humidity, strong winds) that could affect the evaporation rates can be ignored. 

\section{\label{compu}Evaporites Upper Layer Composition Calculation}

   First we build plausible chemical compositions relevant for a lake before its evaporation. Previous calculations made at
thermodynamic equilibrium (D89, C09) correspond to an averaged composition in time and space. Atmospheric precipitation rates are
likely to undergo substantial secular variations --- this is supported by some \emph{Cassini} observations 
\cite[see for instance][]{turtle_eal_2011a}.
In addition, the flux of hydrocarbons falling from the atmosphere likely varies significantly from one 
location to another: in the
equatorial regions the presence of the dune fields indicates an arid climate while lakes, evidences for a wet weather, are located
in the polar regions. These tendencies are confirmed by Global Circulation Models 
\cite[see for instance Figure 4 of C12 based on][]{crespin_etal_2008}. Thus we have considered methane-poor and -rich solvents, 
always containing a \REVsec{few percent of} N$_{2}$.
Concerning the \REVfirst{solids possibly in solution},
we used precipitation rates derived from 1D photochemical models by 
\cite{lavvas_etal_2008a,lavvas_etal_2008b} and \cite{vuitton_etal_2008}--- the precipitation rates are recalled in Table \ref{solids}.
We stress that these rates have to be taken with caution because of the lack of micro-physics in the models. A compound that 
meets an altitude where \REVsec{the temperature equals} its condensation temperature rains out. 
In addition, 3D physical processes
(mainly transport) are ignored, although they can affect significantly the amounts of solid hydrocarbons reaching the surface at
a given place at a particular time.  Two possibilities have been studied.

\begin{itemize}
  \item[\textbullet] In Type 1 solids mixture, where the abundances of these solids have been scaled to atmospheric precipitation rates,
       the most abundant species (\emph{i.e.}, HCN) has its mole fraction set to its value at saturation (ideal solution case). 
       This way, the precipitation \REVfirst{(\emph{i.e.}, saturation)} of solids \REVfirst{in solution} in the liquid begins at the initial time. 
       \REVsec{An assumed initial 
       value below the saturation one, only delaying the starting time of precipitation, keeps the final composition of the upper 
       layer of evaporites unchanged}.

  \item[\textbullet] \REVsec{Type 2} solids mixture, in order to appreciate the effect of evaporation/solids evaporites deposition, we
       constructed also \REVsec{initial mixtures} with uniform 
       \REVsec{abundances, and starting values fixed to}
       the smallest mole fraction at 
       saturation (i.e. that relative to C$_{6}$H$_{6}$). At the evaporation initial
       time, one species begins to precipitate (i.e. C$_{6}$H$_{6}$); the \REVsec{others} saturate latter.
\end{itemize}

  In \REVsec{Table}~\ref{mixturesideal}, results relevant for an ideal solution have been gathered. Two initial mixtures have been considered ---
both ethane rich ($\sim 89\%$ of C$_{2}$H$_{6}$ when evaporation \REVsec{begins}) with $\sim 1\%$ of nitrogen to ensure the liquid physical
state. This abundance of N$_{2}$ is typical of what has been inferred by computations at equilibrium in previous works. 
While $X_{\rm liq}^{\rm ini}$
quantifies the initial chemical composition of the solution, $X_{\rm sol}^{\rm ini}$ represents the abundances of 
\REVfirst{solutes regarded as a single set: $\sum X_{\rm sol}^{\rm ini}= 1$}. The $X_{\rm evap}^{\rm fin}$'s are the 
mole fractions of compounds
finally deposited in the evaporites upper layer. The parameter of enrichment $\Delta$ measures the relative 
enrichment/empoverishment of a given species in the surface evaporites, as compared to the initial composition of solids
\REVfirst{in solution}.

\begin{table}[htbp]
\caption[]{Computed chemical composition ($X_{\rm evap}^{\rm fin}$) of upper layer Titan's lakes drybeds evaporites, in the 
ideal solution hypothesis. Two types of dissolved solids composition have been considered (see text for explanation). 
$X_{i,{\rm sat}}$ is the mole fraction at saturation, $X_{i,{\rm liq}}$ represents the initial composition of the liquid and
$X_{i,{\rm sol}}$ represents the initial abundances of dissolved solids. The \REVsec{$\Delta$'s} show enrichment/empoverishment 
of the resulting upper layer of the evaporite deposit. The assumed temperature is $T= 90$ K. 
The notation $x.y(-n)=x.y \times 10^{-n}$ has been used. \REVfirst{One can notice that ideal solubility of CO$_{2}$ is comparable 
to the values reported by \cite{preston_prausnitz_1970} and \cite{preston_etal_1971} for somewhat higher temperatures around
130-140 K.}}
\begin{center}
{\small
\begin{tabular}{|lc|c|c|c|r|}
\hline
\hline
\multicolumn{6}{l}{Ideal solution} \\
\hline 
Species         & \xsat        &\multicolumn{4}{c|}{Mixture type 1}\\
                & (ideal)      &              &              &              &           \\
                &              & \xliq        & \xsol        & \xevap       & $\Delta$  \\
\hline
CH$_{4}$        &    --        & 10.018\%     & --           &  --          & --        \\
C$_{2}$H$_{6}$  &    --        & 88.804\%     & --           &  --          & --        \\
N$_{2}$         &    --        &  1.002\%     & --           &  --          & --        \\
\hline 
HCN             & $6.46\,(-4)$ & $6.46\,(-4)$ & $3.65\,(-1)$ & $3.82\,(-3)$ & -99 \% \\
C$_{4}$H$_{10}$ & $1.22\,(-1)$ & $5.93\,(-4)$ & $3.35\,(-1)$ & $6.48\,(-1)$ & +94 \% \\
C$_{2}$H$_{2}$  & $5.40\,(-2)$ & $3.62\,(-4)$ & $2.05\,(-1)$ & $3.20\,(-1)$ & +56 \% \\
CH$_{3}$CN      & $3.73\,(-3)$ & $5.42\,(-5)$ & $3.06\,(-2)$ & $2.21\,(-2)$ & -28 \% \\
CO$_{2}$        & $8.72\,(-4)$ & $3.43\,(-5)$ & $1.94\,(-2)$ & $5.16\,(-3)$ & -73 \% \\
C$_{6}$H$_{6}$  & $2.20\,(-4)$ & $8.06\,(-5)$ & $4.55\,(-2)$ & $1.31\,(-3)$ & -97 \% \\
\hline
\hline
                &              & \multicolumn{4}{c|}{Mixture type 2}\\
                &              & \xliq      & \xsol       & \xevap     & $\Delta$  \\
\hline
CH$_{4}$        &    --        & 10.013\%  & --          &  --         & --        \\
C$_{2}$H$_{6}$  &    --        & 88.853\%  & --          &  --         & --        \\
N$_{2}$         &    --        &  1.001\%  & --          &  --         & --        \\
\hline 
HCN             & $6.46\,(-4)$ & $2.20\,(-4)$ & $1.67\,(-1)$ & $6.86\,(-3)$ & -96  \% \\
C$_{4}$H$_{10}$ & $1.22\,(-1)$ & $2.20\,(-4)$ & $1.67\,(-1)$ & $4.71\,(-1)$ & +183 \% \\
C$_{2}$H$_{2}$  & $5.40\,(-2)$ & $2.20\,(-4)$ & $1.67\,(-1)$ & $4.71\,(-1)$ & +183 \% \\
CH$_{3}$CN      & $3.73\,(-3)$ & $2.20\,(-4)$ & $1.67\,(-1)$ & $3.96\,(-2)$ &  -76 \% \\
CO$_{2}$        & $8.72\,(-4)$ & $2.20\,(-4)$ & $1.67\,(-1)$ & $9.26\,(-3)$ &  -94 \% \\
C$_{6}$H$_{6}$  & $2.20\,(-4)$ & $2.20\,(-4)$ & $1.67\,(-1)$ & $2.34\,(-3)$ &  -99 \% \\
\hline
\hline
\end{tabular}
}
\end{center}
\label{mixturesideal}
\end{table}

  \REVsec{As can} be noticed in Table \ref{mixturesideal}, the only species undergoing an enrichment in the surface evaporites layer,
compared to abundances initially taken into account for the dissolved solids, are butane (C$_{4}$H$_{10}$) and acetylene (C$_{2}$H$_{2}$). 
This behavior can be explained by their high solubilities (i.e. high $X_{i,{\rm sat}}$'s). The higher $X_{i,{\rm sat}}$ is, 
the \REVsec{greater} the quantity of dissolved material is. Consequently the saturation occurs later during the evaporation process. 
If we compare $X_{\rm  fin}^{\rm evap}$ obtained for type 1 and type 2 mixtures of solids, we see that evaporite composition 
(perhaps unsurprisingly) depends on the initial abundances of solutes. Our simulation clearly shows, within the framework of our 
current assumptions, that dissolution in methane/ethane solution, followed by evaporation of the solvent, yields surface evaporite 
compositions with high abundances of the most soluble species.


\begin{table}[htbp]

\caption[]{Type 1 mixtures in the case of a non-ideal solution. A methane poor and a methane rich case are considered.

}

\begin{center}

{\small

\begin{tabular}{|l|c|c|c|r|}

\hline

\hline

\multicolumn{5}{l}{Non-ideal solution} \\

\hline 

Species         & \multicolumn{4}{c|}{Mixture type 1, methane poor}   \\

                &             &             &             &           \\

                &  \xliq      & \xsol       & \xevap      & $\Delta$  \\

\hline

CH$_{4}$        &  10.018\%  & --          &  --         & --        \\

C$_{2}$H$_{6}$  &  88.804\%  & --          &  --         & --        \\

N$_{2}$         &   1.002\%  & --          &  --         & --        \\

\hline 

HCN             &  $6.46\,(-4)$ & $3.65\,(-1)$ & $1.52\,(-4)$ & -100 \% \\

C$_{4}$H$_{10}$ &  $5.93\,(-4)$ & $3.35\,(-1)$ & $5.72\,(-1)$ &  +71 \% \\

C$_{2}$H$_{2}$  &  $3.62\,(-4)$ & $2.05\,(-1)$ & $4.20\,(-1)$ & +105 \% \\

CH$_{3}$CN      &  $5.42\,(-5)$ & $3.06\,(-2)$ & $7.14\,(-4)$ &  -98 \% \\

CO$_{2}$        &  $3.43\,(-5)$ & $1.94\,(-2)$ & $6.53\,(-3)$ &  -66 \% \\

C$_{6}$H$_{6}$  &  $8.06\,(-5)$ & $4.55\,(-2)$ & $6.82\,(-4)$ &  -99 \% \\

\hline

\hline

                & \multicolumn{4}{c|}{Mixture type 1, methane rich} \\

                & \xliq      & \xsol       & \xevap      & $\Delta$ \\

\hline

CH$_{4}$        & 90.160 \%  & --          &  --         & --        \\

C$_{2}$H$_{6}$  &  8.662 \%  & --          &  --         & --        \\

N$_{2}$         &  1.002 \%  & --          &  --         & --        \\

\hline 

HCN             & $6.46\,(-4)$ & $3.65\,(-1)$ & $6.92\,(-5)$ & -100 \% \\

C$_{4}$H$_{10}$ & $5.93\,(-4)$ & $3.35\,(-1)$ & $6.69\,(-1)$ & +100 \% \\

C$_{2}$H$_{2}$  & $3.62\,(-4)$ & $2.05\,(-1)$ & $3.24\,(-1)$ &  +58 \% \\

CH$_{3}$CN      & $5.42\,(-5)$ & $3.06\,(-2)$ & $2.94\,(-4)$ &  -99 \% \\

CO$_{2}$        & $3.43\,(-5)$ & $1.94\,(-2)$ & $6.03\,(-3)$ &  -69 \% \\

C$_{6}$H$_{6}$  & $8.06\,(-5)$ & $4.55\,(-2)$ & $3.65\,(-4)$ &  -99 \% \\

\hline

\hline

\end{tabular}

}

\end{center}

\label{mixturesNONideal1}

\end{table}



\begin{table}[htbp]

\caption[]{Type 2 mixtures in the case of a non-ideal solution. A methane poor and a methane rich case are considered.

}

\begin{center}

{\small

\begin{tabular}{|l|c|c|c|r|}

\hline

\hline

\multicolumn{5}{l}{Non-ideal solution} \\

\hline 

Species         & \multicolumn{4}{c|}{Mixture type 2 methane poor}\\

                &             &             &             &           \\

                &  \xliq      & \xsol       & \xevap      & $\Delta$  \\

\hline

CH$_{4}$        &  10.013 \%  & --          &  --         & --        \\

C$_{2}$H$_{6}$  &  88.853 \%  & --          &  --         & --        \\

N$_{2}$         &   1.001 \%  & --          &  --         & --        \\

\hline 

HCN             &  $2.20\,(-4)$ & $1.67\,(-1)$ & $2.18\,(-4)$ & -100 \% \\

C$_{4}$H$_{10}$ &  $2.20\,(-4)$ & $1.67\,(-1)$ & $4.95\,(-1)$ & +197 \% \\

C$_{2}$H$_{2}$  &  $2.20\,(-4)$ & $1.67\,(-1)$ & $4.95\,(-1)$ & +197 \% \\

CH$_{3}$CN      &  $2.20\,(-4)$ & $1.67\,(-1)$ & $1.04\,(-3)$ &  -99 \% \\

CO$_{2}$        &  $2.20\,(-4)$ & $1.67\,(-1)$ & $8.54\,(-3)$ &  -95 \% \\

C$_{6}$H$_{6}$  &  $2.20\,(-4)$ & $1.67\,(-1)$ & $9.55\,(-4)$ &  -99 \% \\

\hline

\hline

                & \multicolumn{4}{c|}{Mixture type 2, methane rich}\\

                & \xliq        & \xsol        & \xevap       & $\Delta$  \\

\hline

CH$_{4}$        & 90.119 \%    & --           &  --          & --        \\

C$_{2}$H$_{6}$  &  8.747 \%    & --           &  --          & --        \\

N$_{2}$         &  1.001 \%    & --           &  --          & --        \\

\hline 

HCN             & $2.20\,(-4)$ & $1.67\,(-1)$ & $6.97\,(-5)$ & -100 \% \\

C$_{4}$H$_{10}$ & $2.20\,(-4)$ & $1.67\,(-1)$ & $6.68\,(-1)$ & +301 \% \\

C$_{2}$H$_{2}$  & $2.20\,(-4)$ & $1.67\,(-1)$ & $3.25\,(-1)$ & + 95 \% \\

CH$_{3}$CN      & $2.20\,(-4)$ & $1.67\,(-1)$ & $2.97\,(-4)$ & -100 \% \\

CO$_{2}$        & $2.20\,(-4)$ & $1.67\,(-1)$ & $6.03\,(-3)$ &  -96 \% \\

C$_{6}$H$_{6}$  & $2.20\,(-4)$ & $1.67\,(-1)$ & $3.67\,(-4)$ & -100 \% \\

\hline

\hline

\end{tabular}

}

\end{center}

\label{mixturesNONideal2}

\end{table}


   We stress \REVsec{that an identical value of the enrichment $\Delta$} is the consequence of a saturation of solutes that occurs 
at the very end of the evaporation. Of course, a solvent of a different composition (\emph{e.g.}, a methane rich one) leads strictly 
to the same result because here we are making the calculations by adopting the ideal \REVsec{solution} hypothesis.

   The results of non-ideal simulations for the regular solution $\Gamma_{i}$'s have been gathered in Table~\ref{mixturesNONideal1}
and Table~\ref{mixturesNONideal2}.  Tables~\ref{mixturesNONideal1} and \ref{mixturesNONideal2} are respectively devoted to dissolved
solids mixture type 1 and type 2. For each of these types, cases of methane rich and poor solvent are considered. As can be noticed 
in Table~\ref{mixturesNONideal1} and Table~\ref{mixturesNONideal2}, the general trend remains the same: butane and acetylene, if 
present in the initial mixture, are the dominant species in the upper evaporite layer. The difference between the results of methane 
rich and poor are explained by the non-ideality of the solution: in such a situation the molecules undergo interactions. In this 
way, solvents with different compositions are not equivalent.


\begin{table}[htbp]

\caption[]{Influence of the initial nitrogen abundance on final evaporite layer composition.

$T= 90$ K, non-ideal. Methane rich solvent ($X^{ini}_{\rm CH_{4}} \simeq 90$\%)}

\begin{center}

{\small

\begin{tabular}{lccc}

\hline

$X^{\rm ini}_{\rm N_{2}}$ &   0.5 \%    &   1 \%      &   3 \%     \\

\hline

                          &   $\Delta$  &   $\Delta$  &   $\Delta$ \\

HCN                       &   -99.9 \%  &  -100.0 \%  &  -100.0 \% \\ 

C$_{4}$H$_{10}$           &   +270.5 \%  &  +300.9 \%  &  +377.5 \% \\

C$_{2}$H$_{2}$            &   +125.0 \%  &  +95.1 \%  &   +19.1 \% \\

CH$_{3}$CN                &   -99.7 \%  &   -99.8 \%  &  -100.0 \% \\

CO$_{2}$                  &   -96.2 \%  &   -96.4 \%  &   -96.7 \% \\

C$_{6}$H$_{6}$            &   -99.7 \%  &   -99.8 \%  &  -100.0 \% \\

\hline

\end{tabular}

}

\end{center}

\label{influenceN2}

\end{table}


   The \REVsec{content of nitrogen} is expected to vary slightly for different bodies of liquid. Hence we test for sensitivity of 
evaporite composition regarding the abundances of nitrogen in the solvent. Table~\ref{influenceN2} shows the enrichment parameter 
$\Delta$ 
\REVfirst{for all solutes}
in three cases: $X^{\rm ini}_{\rm N_{2}}= 0.5$\%, 1\% and 3\%. All of these computations have been made for a methane rich solvent 
for a temperature of 90 K. Clearly, solvents with a high nitrogen abundance appear to favor C$_{4}$H$_{10}$ as a main constituent of
surface evaporites.


\begin{table}[htbp]

\caption[]{Influence of the temperature on final evaporite layer composition. The initial nitrogen mole fraction

is fixed to $X_{\rm N_{2}}^{\rm ini}= 1$\%, the solution is a non-ideal one, and type 1 for the initial dissolved solids mixture

has been considered.}

\begin{center}

{\small

\begin{tabular}{lcccccc}

\hline

\hline

Species         & \multicolumn{2}{c}{$T= 85$ K}  &  \multicolumn{2}{c}{$T= 90$ K}  & \multicolumn{2}{c}{$T= 95$ K}  \\

\multicolumn{7}{c}{methane poor solvent}\\

\hline 

                & \xevap       &  $\Delta$       &  \xevap       &  $\Delta$       & \xevap       &  $\Delta$       \\

HCN             & $1.17\,(-4)$ &  -100           &  $1.52\,(-4)$ &  -100           & $1.84\,(-4)$ &  -100           \\

C$_{4}$H$_{10}$ & $5.19\,(-1)$ &  +54.9         &  $5.72\,(-1)$ &   +70.7         & $6.20\,(-1)$ &   +85.0         \\

C$_{2}$H$_{2}$  & $4.76\,(-1)$ &  +132.4         &  $4.20\,(-1)$ &  +105.3         & $3.70\,(-1)$ &  +80.8         \\

CH$_{3}$CN      & $6.52\,(-4)$ &   -97.9         &  $7.14\,(-4)$ &   -97.7         & $7.39\,(-4)$ &   -97.6         \\

CO$_{2}$        & $4.59\,(-3)$ &   -76.3         &  $6.53\,(-3)$ &   -66.3         & $8.77\,(-3)$ &   -54.7         \\

C$_{6}$H$_{6}$  & $5.54\,(-4)$ &   -98.8         &  $6.82\,(-4)$ &   -98.5         & $7.88\,(-4)$ &   -98.3         \\

\hline

\multicolumn{7}{c}{methane rich solvent}\\

\hline 

                & \xevap       &  $\Delta$       &  \xevap       &  $\Delta$       & \xevap       &  $\Delta$       \\

HCN             & $5.18\,(-5)$ &  -100           &  $6.92\,(-5)$ &  -100           & $8.76\,(-5)$ &   -100          \\

C$_{4}$H$_{10}$ & $6.31\,(-1)$ &  +88.3         &  $6.69\,(-1)$ &   +99.8         & $7.01\,(-1)$ &    +109.4        \\

C$_{2}$H$_{2}$  & $3.65\,(-1)$ &  +78.1         &  $3.24\,(-1)$ &   +58.4         & $2.90\,(-1)$ &    +41.6        \\

CH$_{3}$CN      & $2.62\,(-4)$ &   -99.2         &  $2.94\,(-4)$ &   -99.0         & $3.17\,(-4)$ &    -99.0        \\

CO$_{2}$        & $4.29\,(-3)$ &   -77.9         &  $6.03\,(-3)$ &   -68.9         & $8.07\,(-3)$ &    -58.3        \\

C$_{6}$H$_{6}$  & $2.96\,(-4)$ &   -99.4         &  $3.65\,(-4)$ &   -99.2         & $4.28\,(-4)$ &    -99.1        \\

\hline

\hline

\end{tabular}

}

\end{center}

\label{influT}

\end{table}


 The temperature also influences the solubility of solids. For a range of temperatures (\emph{i.e.}, 85 K, 90 K, and 95 K), the 
computed final compositions have been reported in Table~\ref{influT}. High values of temperature favors high-butane evaporite content, 
while acetylene is disfavored at high temperature. However the major tendency, the prominence of butane and acetylene, is robust.\\
 Finally, we stress that during all our calculations, for each time, we have checked that the density of species precipitating 
remained lower than the liquid solution value. Then, according to the simulations performed in this work, the organic precipitated solids
would never float at the surface of the Titan's lakes 
\citep[at least not without unusual circumstances such as those suggested by][]{hofgartner_lunine_2013}.


\section{\label{discuss}Discussion and Conclusion}

   As it has been shown in the previous section, the 
\REVsec{composition obtained}
for the superficial layer of evaporite is the result of the influence of two main factors: (1) the initial composition of dissolved 
solids, and (2) the mole fraction at saturation values of the solids, provided by Equation (\ref{Xs}). For a given initial composition, 
species with the highest $X_{i, {\rm sat}}$ (\emph{i.e.}, the lowest energy of cohesion) remain dissolved  for a longer time in the 
solvent during the evaporation and finally become the major \REVsec{constituents} of the last layer of deposits. The value of 
\REVsec{the} melting temperature $T_{\rm m}$ and enthalpy of melting determine the concentration at saturation. We furthermore 
emphasize that, due to the special non-linear scale of time used in this work, the depth of different layers of evaporite cannot be 
computed. In a same way, the composition of possible evaporite annuli around a lake \citep[see for instance][]{hayes_etal_2010} could 
not be estimated. For a given initial composition of \REVfirst{solids in solution}, we can only compute the composition of the 
external surface of the evaporite deposit.\\



\bibliographystyle{elsarticle-harv}

\vspace{1cm} We acknowledge financial support from the Observatoire des Sciences de l'Univers THETA Franche-Comté-Bourgogne, France.
\variouscom{We thank Sandrine Vinatier and Panayotis Lavvas for scientific discussion; and the anonymous Reviewers who improved 
the clarity of the paper with their remarks and comments. JWB is funded by the NASA CDAPS Program Grant \#NNX12AC28G.}

\end{document}